\documentstyle[psbox]{WORKSHOP}
\begin{document}

\newcommand{\xte}{{\textit{RXTE}}}
\newcommand{\rosat}{{\textit{ROSAT}}}
\newcommand{\g}{$\gamma$}

\title{Orbital and superorbital variability and their coupling in X-ray binaries}

\author{
Andrzej A. Zdziarski,$^1$ Juri Poutanen,$^2$ Askar Ibragimov,$^{2,3}$ Marek Gierli{\'n}ski,$^{4,5}$ and Linqing Wen$^{6,7,8}$\\[12pt]
$^1$Centrum Astronomiczne im.\ M. Kopernika, Bartycka 18, 00-716 Warszawa, Poland\\
$^2$Astronomy Division, Department of Physical Sciences, PO Box 3000, FIN-90014 University of Oulu, Finland\\
$^3$Kazan State University, Astronomy Department, Kremlyovskaya 18, 420008 Kazan, Russia \\
$^4$Department of Physics, University of Durham, Durham DH1~3LE, UK\\
$^5$Obserwatorium Astronomiczne Uniwersytetu Jagiello\'nskiego, Orla 171, 30-244 Krak{\'o}w, Poland\\
$^6$Max-Planck-Institut f{\"u}r Gravitationsphysik, Albert-Einstein-Institut, Am M{\"u}hlenberg 1, D-14476 Golm, Germany\\
$^7$Division of Physics, Mathematics and Astronomy, Caltech, Pasadena, CA 91125, USA\\
$^8$School of Physics, University of Western Australia, Crawley, WA 6009, Australia\\
{\it E-mail (AAZ): aaz@camk.edu.pl} 
}

\abst{
We review X-ray flux modulation from X-ray binaries on time scales corresponding to the orbital period and those at longer time scales (so called superorbital). Those modulations provide a powerful tool to constrain geometry of the accretion flow. The most common cause of the superorbital variability appears to be precession. We then discuss two specific examples of discoveries of a coupling between the two types of variability and their physical interpretation. One is Cyg X-1, a black-hole binary with a high-mass companion, in which case we find the presence of an accretion bulge formed by collision of the stellar wind with the outer edge of the precessing accretion disc. The other is 4U 1820--303, a neutron star accreting from a low-mass white dwarf, in which case we interpret the superorbital variability as accretion rate modulation induced by interactions in a triple stellar system. Then, the varying accretion rate leads to changes of the size of the accretion bulge in that system, obscuring the centrally-emitted X-rays.
}

\kword{accretion, accretion discs --- stars: individual: (Cyg X-1, HDE 226868, 4U 1820--30) --- X-rays: binaries --- X-rays: stars}

\maketitle
\thispagestyle{empty}

\section{Introduction}
\label{intro}

A large number of X-ray binaries show X-ray flux periodicities at their orbital period. The current comprehensive list of such objects is given by Wen et al.\ (2006). Using strict criteria for the statistical significance of the presence of a modulation, they find 33 secure orbital periods in the \xte/ASM data. Due to the limitation of the ASM sampling, they were unable to detect the 685-s X-ray binary period of the ultracompact low-mass X-ray binary (LMXB) 4U 1820--303, which, however, is clearly detected, e.g., in \xte/PCA data (e.g., Zdziarski et al.\ 2007b, hereafter Z07, and references therein). Even faster X-ray modulations, at 569 s and 321 s, are found in \rosat\/ data for RX J1914.4+2456 (Cropper et al.\ 1998) and RX J0806.3+1527 (Israel et al.\ 1999), respectively. These two systems contain pairs of white dwarfs orbiting each other at the above respective period (Cropper et al.\ 1998; Israel et al.\ 2002; Ramsay et al.\ 2002). 

In addition, a number of X-ray binaries show X-ray modulations at periods longer than their orbital periods, so-called superorbital periodicity. Sood et al.\ (2007) state that 35 such systems have been reported in literature. However, many of those claims remain uncertain, and Sood et al.\ (2007) selected 19 of them as secure. However, even some of those appear, in our opinion, uncertain. For example, they list the high-mass X-ray binary (HMXB) Cen X-3 as having a superorbital period of 140 d. On the other hand, the same object is presented by Raichur \& Paul (2008) as a case of the {\it lack\/} of stable superorbital modulation. Another example is 4U 1916--053, which is listed as having a 199-d period in Sood et al.\ (2007) whereas Homer et al.\ (2001) (and later Wen et al.\ 2006) find no trace of such periodicity. Wen et al.\ (2006), using a uniform analysis method, find 6 objects as having coherent superorbital modulations in the \xte/ASM data, and further 5 with quasi-periodicities. Using their method with no pre-selection of data, they did not find a superorbital periodicity in the black-hole HMXB Cyg X-1. However, its $\sim$150-d period manifests itself mostly in the hard spectral state, where its statistical significance is very high (e.g., Brocksopp et al.\ 1999a; Karitskaya et al.\ 2001; {\"O}zdemir \& Demircan 2001; Lachowicz et al. 2006; Ibragimov et al.\ 2007; Poutanen et al.\ 2008, hereafter PZI08). Another black-hole binary in which the superorbital modulation manifests itself only in the hard state appears to be the LMXB GX 339--4, in which it shows a $\sim$220-d periodicity, whereas that modulation is absent in the soft state (Zdziarski et al.\ 2004). (See, e.g., Zdziarski \& Gierli\'nski 2004; Done et al.\ 2007 for reviews and physical interpretation of the spectral states of black-hole binaries.)

The MAXI, an X-ray all-sky monitor mission on the Japanese Experiment Module of the {\it International Space Station}, is scheduled to be launched in 2009 March. Due to its unprecedented sensitivity of a few mCrab in a day covering most of the sky, the MAXI will monitor variability of a large number of X-ray sources at much lower flux levels than is now possible with the \xte/ASM. It will thus be able to greatly enhance our knowledge of the variability of X-ray binaries, in particular of their orbital and superorbital periodicities. 

\section{Physical causes of orbital and superorbital flux periodicities}
\label{physical}

An orbital periodicity may be caused by a number of effects. Here, we review also some effects affecting other wavelengths than X-rays. First, a source associated with a compact object in a binary (usually a HMXB) may be eclipsed by its companion, see a list in Wen et al.\ (2006). Second, a flux modulation may be caused by an optically-thick disc rim (which is highest at the point of impact of the gas stream from the inner Lagrangian point in case of a donor filling its Roche lobe), obscuring the disc and/or its corona (e.g., White \& Swank 1982; Hellier \& Mason 1989; Heinz \& Nowak 2001), usually in LMXBs. This obscuration may lead to strong partial eclipses in so-called X-ray dippers. More generally, the disc and any associated structures may depart from its axial symmetry due to the influence of the companion, which may cause an orbital modulation. Third, wind from a high-mass companion may absorb/scatter the emission from the vicinity of the compact object, and the degree of absorption will depend on the orbital phase. In the case of Cyg X-1, both X-ray and radio emission are modulated by this effect, which modulations were modeled by, e.g., Wen et al.\ (1999) and Szostek \& Zdziarski (2007), respectively. Fourth, phase-dependent absorption (via photon-photon pair production) of high-energy \g-rays may occur in a photon field axially asymmetric with respect to the compact object, especially that of the stellar photons (e.g., Bednarek 1997; Dubus 2006). A fifth effect of the companion is reflection or reprocessing of the emission from around the compact object on the surface of the companion facing the compact object (Basko 1978). This effects appears to be responsible for, e.g., the UV flux modulation from the X-ray binary 4U 1820--303 (Arons \& King 1993; Anderson et al.\ 1997). Finally, the optical/UV emission of the companion will be modulated if its shape departs from the spherical symmetry by partially or fully filling its Roche lobe, which effect is seen in Cyg X-1, e.g., Brocksopp et al.\ (1999b).

Then, there will be an intrinsic dependence of the emitted flux on the orbital phase if the orbit is elliptical. This leads, e.g., to periodic outbursts around the periastron of Cir X-1 (e.g., Parkinson et al.\ 2003) and Be/X-ray binaries (see, e.g., Coe 2000; Negueruela 2004 for reviews) in X-rays, and sometimes, at other wavelengths.  Also, some orbital flux modulation may be due to the Doppler effect, which is in principle observable (Postnov \& Shakura 1987), but has not yet been detected in a binary. (Obviously, the Doppler effect leads to widely observed shifts of spectral lines from binaries.)

The observed superorbital variability appears in most cases compatible with being caused by accretion disc and/or jet precession, which either results in variable obscuration of emitted X-rays as in Her X-1 (Katz 1973), or changes the viewing angle of the presumed anisotropic emitter, as in SS 433 (Katz 1980) or Cyg X-1 (e.g., Lachowicz et al.\ 2006; Ibragimov et al.\ 2007; PZI08), or both. The precession may be induced via tidal forces by the companion (Larwood 1998), with additional forces exerted by irradiation of the disc by the central X-ray source (e.g., Ogilvie \& Dubus 2001). The only certainly known exception, in which the superorbital periodicity is clearly caused by modulation of the accretion rate, $\dot M$ (and thus not by a changing viewing angle of the source), appears to be 4U 1820--303 (Zdziarski et al.\ 2007a; Z07, and references therein). The $\dot M$ modulation can be achieved in a triple system via a secular modulation of the eccentricity of the inner system, proposed for this source by Chou \& Grindlay (2001), and calculated in detail by Zdziarski et al.\ (2007a). Another system possibly of this type is the HMXB 2S 0114+650, where Farrell et al.\ (2008) have argued for changes of the $\dot M$ based on the spectral variability of the system.

\section{Coupling between orbital and superorbital modulations}
\label{coupling}

\begin{figure}[t]
\centering
\psbox[xsize=6.5cm]{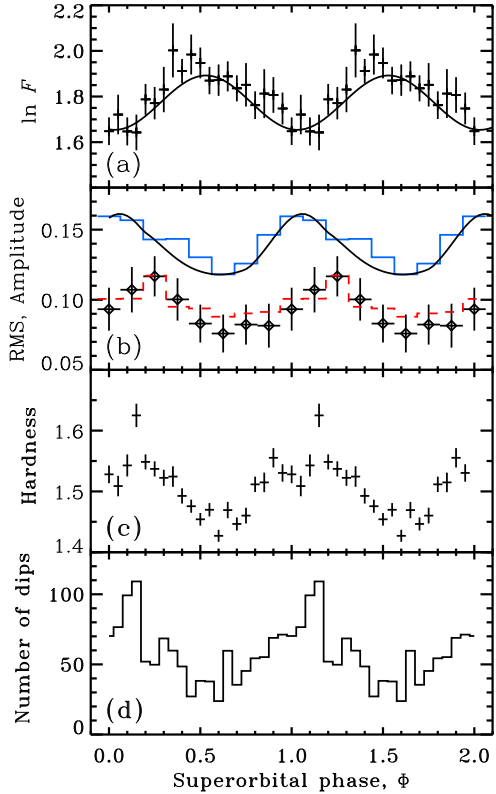}
\caption{ 
(a) The superorbital phase diagram for the ASM 1.5--3 keV channel (the unit of $F$ is count s$^{-1}$). (b) Comparison of the characterization of the 1.5--3 keV orbital modulation dependence on the superorbital phase using different methods. The crosses show the intrinsic fractional rms of the orbital modulation.  The solid histogram gives the orbital variability amplitude fitted by sum of three harmonics. The dashed histogram gives the corresponding rms for the fitting function. The solid curves in (a--b) show the dependencies for a theoretical outflow model fitted to the data by PZI08. (c) The mean hardness ratio (5--12 keV/1.5--3 keV), and (d) the distribution of X-ray dips in the
hard state. A visible offset of the ephemeris with respect to $\Phi=0$ is due to the use of an ephemeris based also on older data (Lachowicz et al.\ 2006). 
}
\label{rms_asm}
\end{figure}

A number of binaries show both orbital and superorbital modulations. Those currently known are LMC X-4, 2S 0114+650, SMC X-1, Her X-1, SS 433, 4U 1820--303 and Cyg X-1. An interesting issue then is whether there is any dependence of the parameters of the orbital modulation on the superorbital phase (or, similarly, on an average of the flux level). The shape of the profile of the orbital modulation in Her X-1 was found to depend on its superorbital phase (Scott \& Leahy 1999), which appears to be due to the shadowing effect of the precessing accretion disc and scattering in its wind in that system. Analogous dependencies of the shape of the orbital modulation on the average flux level have been found (Raichur \& Paul 2008) in LMC X-4, SMC X-1, Her X-1, as well as in Cen X-3 (which object, however, shows a chaotic superorbital behavior). 

The Be/X-ray binary LS I +61$^\circ$303 shows orbital variability in the radio, X-ray and TeV emission, and a superorbital variability of the peak radio flux during an orbit (Gregory et al.\ 1999; Gregory 2002). Gregory (2002) found a marked dependence of the phase of the peak of the orbital radio modulation on the superorbital phase. The presence of such a dependence may be due to interaction of the pulsar in that system with a variable circumstellar Be decretion disc (Gregory 2002; Zdziarski et al.\ 2008), but details remain unknown.

Then, strong coupling between the orbital and superorbital modulations is seen in the HMXB Cyg X-1 (PZI08) and in the LMXB 4U 1820--303 (Z07), which consists of a white dwarf accreting onto a weakly-magnetized neutron star. We review these two objects in detail below in Sections \ref{cygx1}--\ref{4u1820} 

\section{The dependence of the orbital modulation on the superorbital phase in Cyg X-1}
\label{cygx1}

\begin{figure}[t]
\centering
\psbox[xsize=6.3cm]{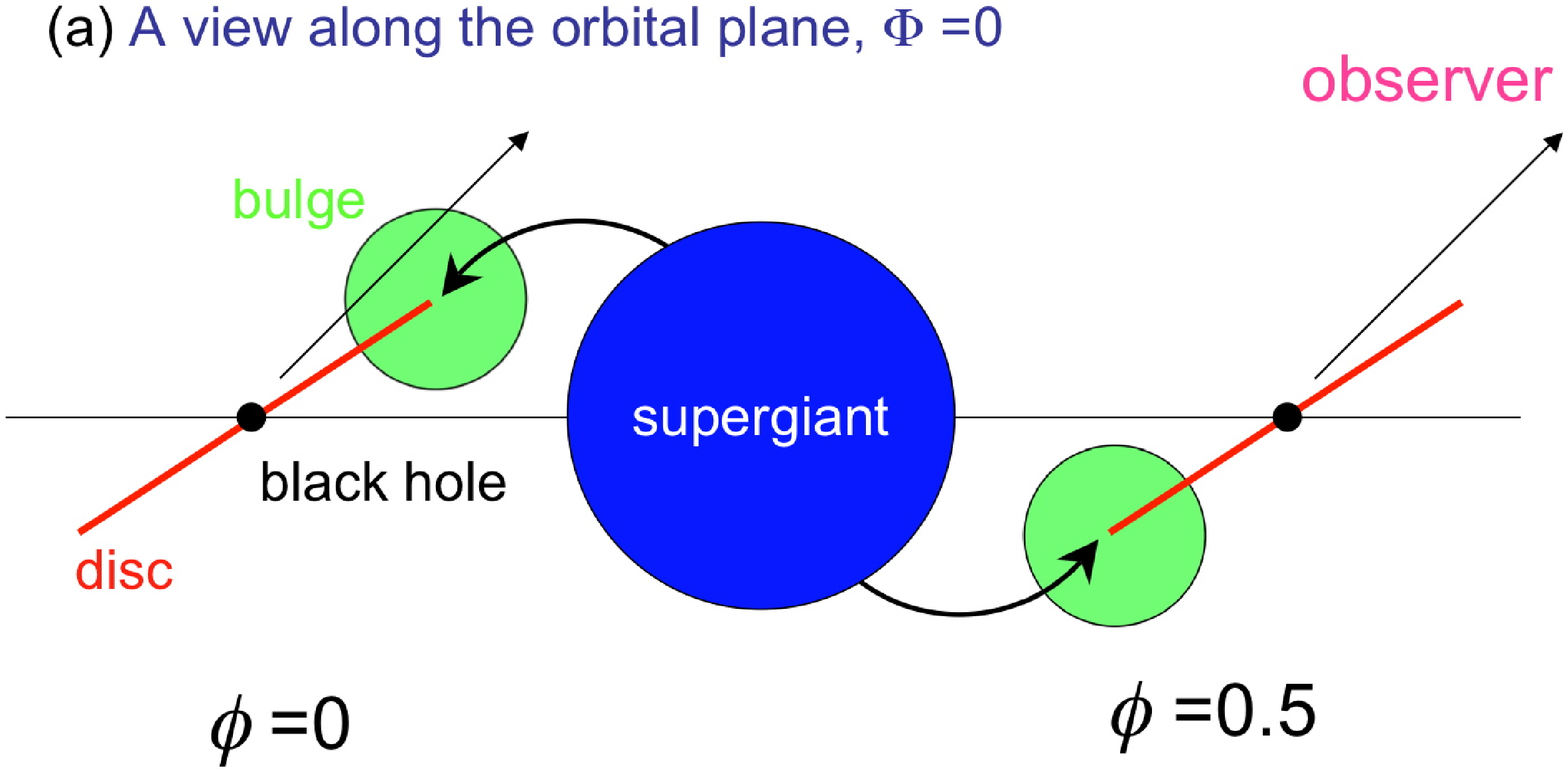}
\psbox[xsize=6.3cm]{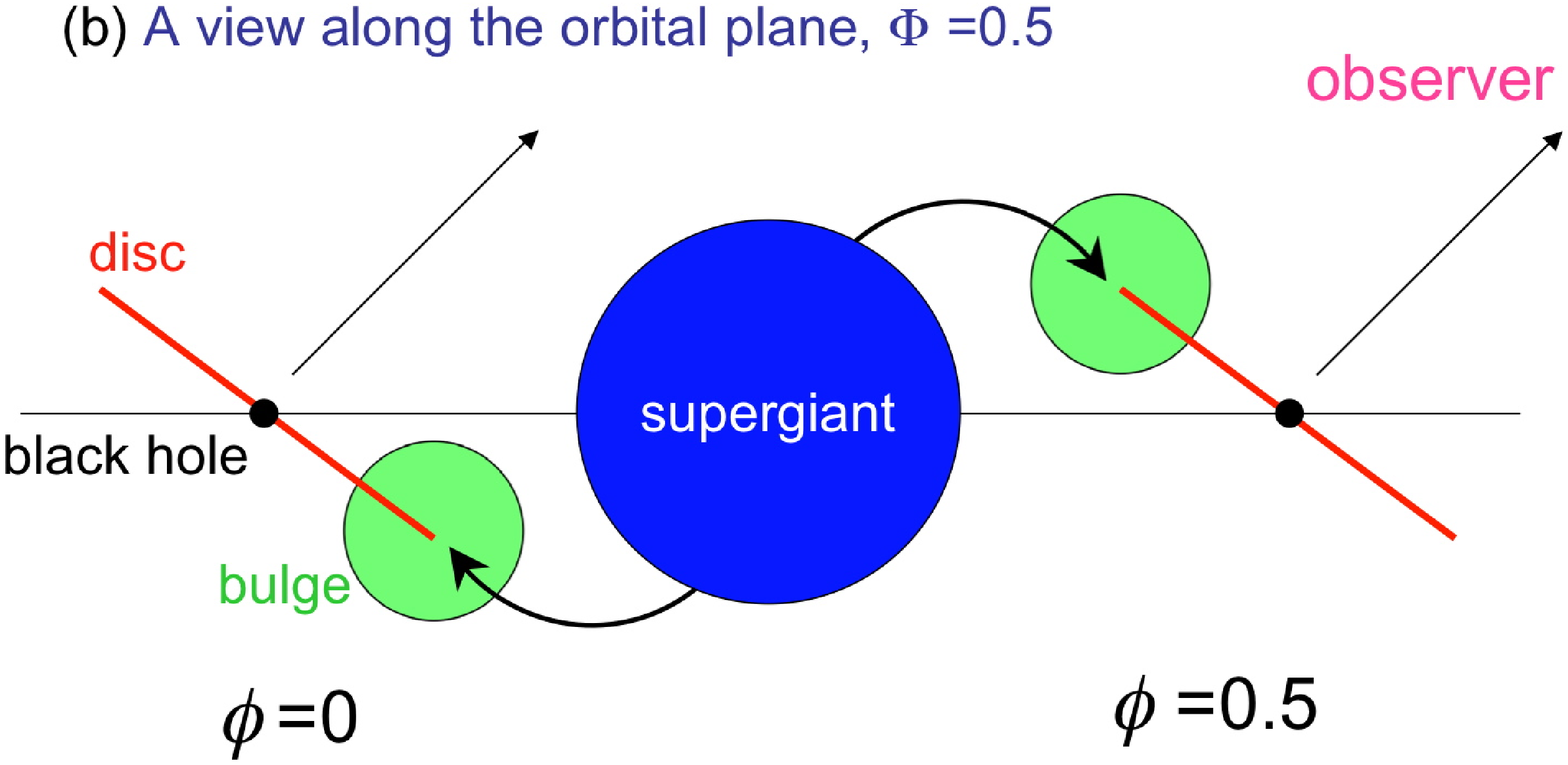}
\psbox[xsize=6.3cm]{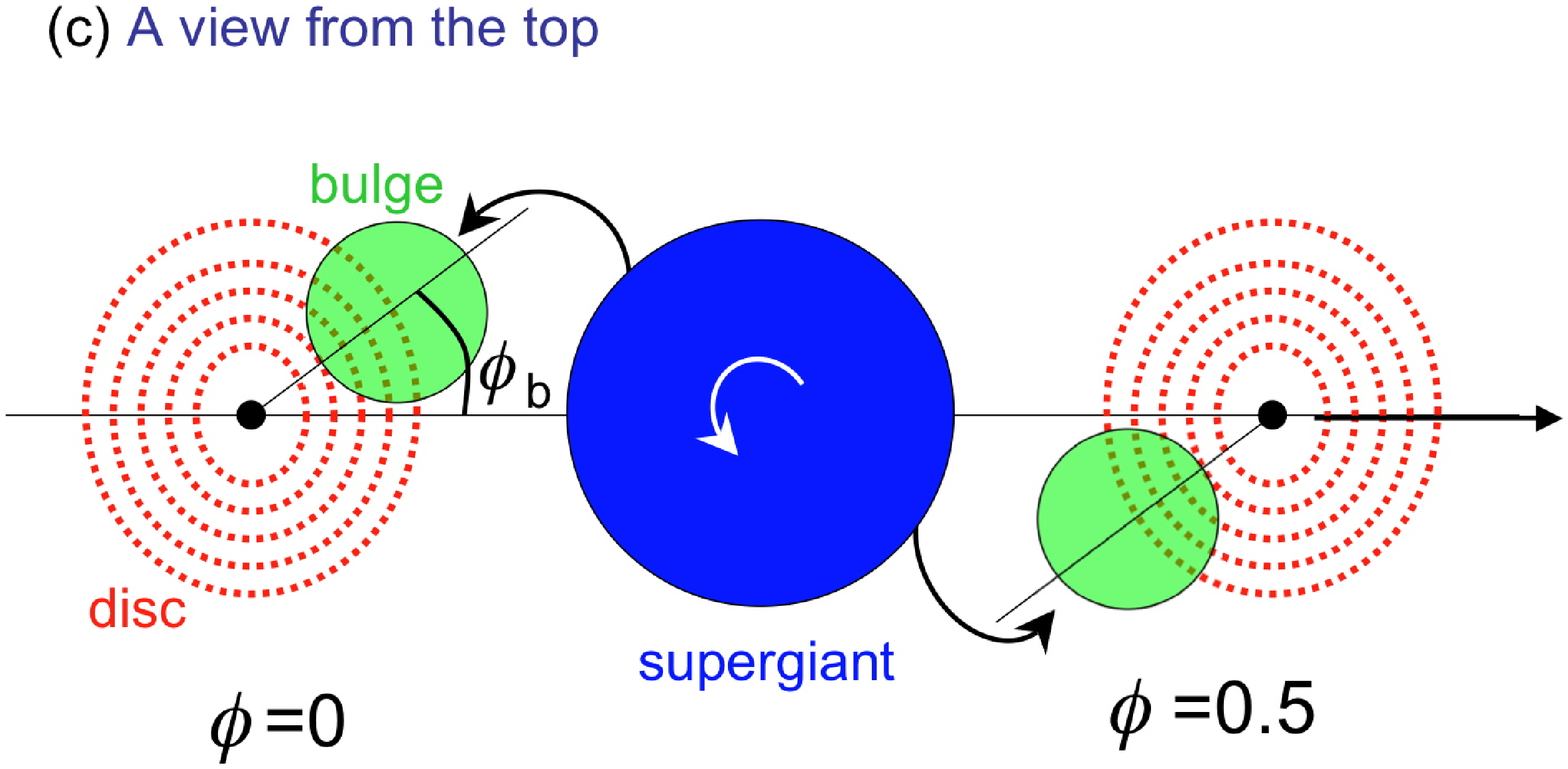}
\caption{A bulge at the outer edge of a precessing inclined disc. The material in the bulge absorbs some of the X-ray emission originating close to the disc center. The orbital modulation due to the bulge is seen to strongly depend on the superorbital phase. (Orbital modulation due to the stellar wind itself is not illustrated here for clarity.) A view along the orbital plane: (a) the superorbital phase $\Phi= 0$, when the disc is seen closest to edge-on and the effect of the bulge is strongest; (b) the opposite case of the superorbital phase $\Phi= 0.5$. (c) A view from the top.
}
\label{model}
\end{figure}

\begin{figure*}[t]
\centering
\psbox[xsize=10.cm]{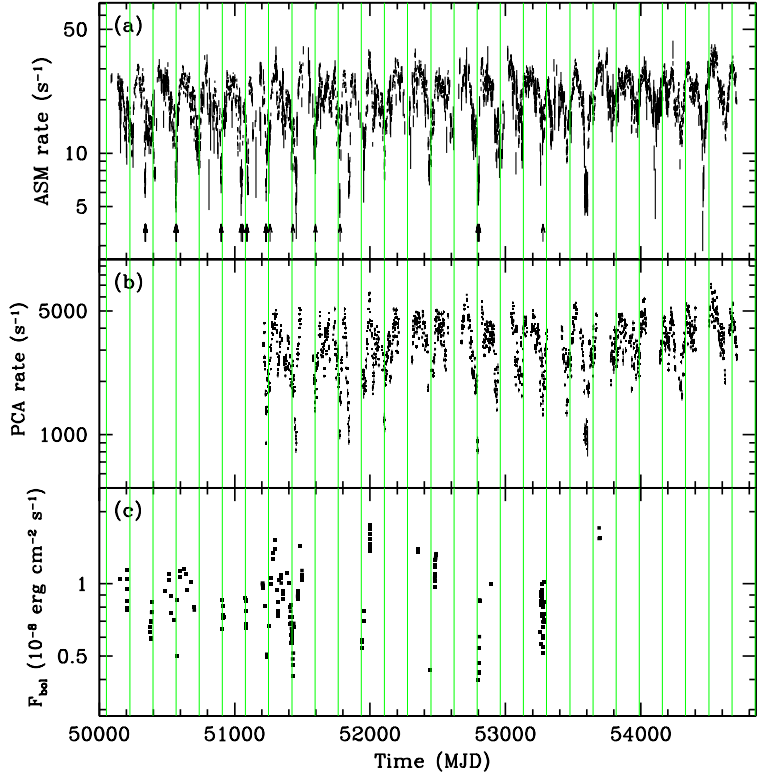}
\caption{(a) The \xte/ASM light curve of 4U 1820--303. The short arrows shows the times of detected X-ray bursts. (b) The PCA light curve based on Galactic Bulge scans. (c) The light curve of the bolometric flux calculated from the PCA/HEXTE data. See Z07 for details. The ASM and PCA light curves have been extended up to MJD $\sim$54700. The vertical lines show the minima (the phase $\Phi=0$) of the superorbital cycle according to the ephemeris of Chou \& Grindlay (2001), which ephemeris is seen to become increasingly inaccurate in recent epochs.
\label{lc} }
\end{figure*}

Such a dependence was searched for, studied, and interpreted theoretically by PZI08. They used the \xte/ASM data, in which both orbital and superorbital modulations are clearly seen, see Section \ref{intro} The physical cause of the orbital X-ray modulation is bound-free absorption in the stellar wind, whereas the widely accepted underlying cause of the superorbital, $\sim$150-d, modulation is accretion disc precession, see Section \ref{physical} As found by Ibragimov et al.\ (2007), the resulting superorbital flux variability is most likely due to an anisotropy of the X-ray emission from the accretion flow (rather than due to absorption, obscuration or scattering). 

PZI08 found the fractional amplitude of the orbital modulation strongly depends on the superorbital phase in the hard state, see Fig.\ \ref{rms_asm}a, b. Namely, the orbital modulation is strongest when the average flux level is lowest. The effect is visible in all of the three ASM channels, but it is strongest in the lowest-energy channel, 1.5--3 keV, where the orbital modulation amplitude is highest (which reflects the energy dependence of the bound-free cross section averaged over the cosmic abundances).

A theoretical interpretation of this effect proposed by PZI08 is as follows. The cause of the part of the orbital modulation dependent on the superorbital phase is a structure at the outer edge of the accretion disc on the side of the companion. Such a structure is expected theoretically. Formation of the accretion disc in Cyg X-1 by a focused wind (Gies \& Bolton 1986) leads, most likely, to a condensation of the wind matter near the disc outer edge on the side of the companion in the form of a bulge, similar to the disc bulge inferred to be present in low-mass X-ray binaries, e.g., White \& Holt 1982; White \& Swank 1982; Parmar \& White 1988). On the other hand, the bulge can also be formed (see, e.g., Boroson et al.\ 2001) by a shock wave in the wind when it encounters the gravity of the companion, the disc, or a wind from the disc, which is also likely to be present. In any case, when the fast, $>1000$ km s$^{-1}$, wind is stopped, the density increases dramatically.

Such a bulge is shown in Fig.\ \ref{model}. We see that at the minimum of the superorbital cycle (i.e., at the superorbital phase of $\Phi\simeq 0$), when the disc is most edge-on (Fig.\ \ref{model}a), the bulge crosses the line of sight to the X-ray source (which is close to the black hole) around the superior conjunction, i.e., at the orbital phase of $\phi\simeq 0$, and it is away from the line of sight at the inferior conjunction, i.e., at the orbital phase of $\phi\simeq 0.5$. On the other hand, at the maximum of the superorbital cycle  ($\Phi\simeq 0.5$), the disc is most face-on (Fig.\ \ref{model}b), and the bulge is away from the line of sight at any orbital phase. Thus, the bulge introduces additional absorption only around $\Phi\simeq 0.5$, just as observed. 

This model was put into the form of a set of equations and then fitted to the data by PZI08. They considered a number of models for the anisotropy of the X-ray emission, leading to the superorbital modulation, following Ibragimov et al.\ (2007). An exponential absorption profile of the bulge and either an isotropic or focused wind were assumed. In all models, the total optical depth of the bulge measured from its center was found to be $\simeq $1. The bulge center was found to be displaced from the line connecting the centers of the stars by $\simeq 25^\circ$. The tilt of the precessing disc was found to be compatible with $\sim 10^\circ$, at the assumed inclination of $40^\circ$. Prograde precession was strongly favored by the fits. 

PZI08 also have shown that both the X-ray spectral hardness and the frequency of X-ray dips (caused by absorption by wind blobs, Ba{\l}uci\'nska-Church et al.\ 2000) increase towards the superorbital phase of 0, see Fig.\ \ref{rms_asm}c, d, which is also explained by the above model. Furthermore, the coupling of the orbital and superorbital modulations leads to appearance of asymmetric beat frequencies in the power spectrum, explaining the finding of such asymmetry by Lachowicz et al.\ (2006). 

\section{The dependence of the orbital modulation on the spectral state in 4U 1820--303}
\label{4u1820}

Such a dependence was searched for, studied, and interpreted theoretically by Z07. Long-term light curves of the system are shown in Fig.\ \ref{lc}. We see a very pronounced superorbital variability, with the period of $\simeq$170 d. The strongest argument for the intrinsic, $\dot M$-driven, character of this variability is the occurrence of X-ray bursts (which generally take place at low accretion rates) only during the deep minima of the light curve, see Fig.\ \ref{lc}a. As proposed by Chou \& Grindlay (2001), the physical cause of the $\dot M$ modulation is quasi-periodically varying eccentricity, $e$, of the binary induced by a third star in the system, via the so-called Kozai effect (Kozai 1962). This was calculated in detail by Zdziarski et al.\ (2007a), who have modeled such $e$ evolution and calculated the resulting $\dot M$ variability in a Roche-lobe overflow model.  

\begin{figure}[t]
\centering
\psbox[xsize=7.4cm]{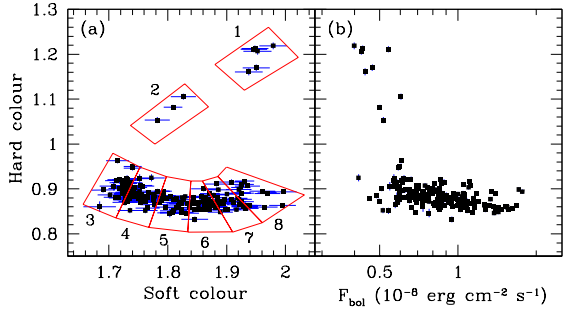}
\caption{(a) The intrinsic colour-colour diagram, as given by the ratio of the energy fluxes in the photon energy ranges of (9.7--16 keV)/(6.4--9.7 keV) and (4.0--6.4 keV)/(3.0--4.0 keV) for the hard colour and soft colour, respectively. 
The boxes show a division into 8 spectral substates. (b) The corresponding dependence of the hard colour on the bolometric flux. From Z07.
\label{col_col} }
\end{figure}

\begin{figure}
\centering
\psbox[xsize=6.9cm]{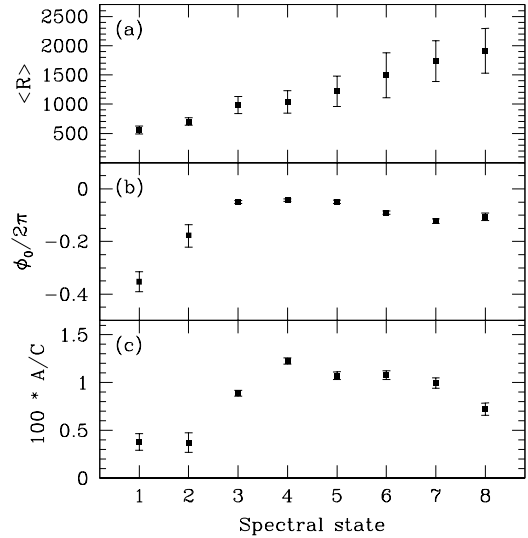}
\caption{ The average PCA count rate ($\langle R\rangle$), the offset orbital phase ($\phi_0$), and the fractional amplitude of the orbital modulation (denoted as $A/C$) vs.\ the spectral state, defined in Fig.\ \ref{col_col}. The error bars on $\langle R\rangle$ represent the standard deviation, not the errors of the average (which are negligibly small). From Z07.
\label{fit_state} }
\end{figure}

The orbital modulation of the system was discovered in X-rays (Stella et al.\ 1987). Given the very short orbital period, it is not detectable in the ASM data, and therefore Z07 used PCA/HEXTE observations of 4U 1820--303. Their light curve is shown in Fig.\ \ref{lc}c. We see their coverage of the superorbital variability is rather sparse, and it has proven difficult to study the orbital modulation directly as a function of the superorbital phase. On the other hand, the superorbital cycle is closely related to the spectral variability of the system, going between the so-called atoll (at low $\dot M$) and banana (at high $\dot M$) spectral states (Bloser et al.\ 2000). Therefore, Z07 have divided the pointed X-ray observations into 8 spectral substates, with the $\dot M$ increasing from the state 1 to 8, as shown in Fig.\ \ref{col_col}.

Then the orbital modulation was fitted within each substate by a sinusoidal dependence. Fig.\ \ref{fit_state} shows the results. We see that the PCA count rate increases with the state number (as expected). The fractional modulation amplitude very significantly increases from the atoll (1--2) to banana (3--8) states. Also, the offset phase increases. Z07 interpreted the results in terms of the size and location of the bulge at the disc rim (partially obscuring the central source) changing with the variable accretion rate. Z07 also found that the fractional modulation was independent of the photon energy. This argues for the part of the bulge obscuring the X-rays being almost completely ionized. More details and discussion of other options are given in Z07.

\section{Discussion}

We have reviewed orbital and superorbital modulation of X-ray binaries, concentrating on the X-ray spectral band. The physical causes of the orbital modulation are relatively well understood, see Section \ref{physical} The most common cause of the superorbital modulation appears be precession of accretion disc. However, details remain unclear, in particular what drives the precession, whether it is prograde or retrograde, and how a viscous disc can coherently precess. On the other hand, some objects, especially 4U 1820--303, show very strong evidence of the superorbital variability being due to periodic changes of $\dot M$, and not due to precession. However, physical details also here remain unclear. One possibility is the effect of a third star in a hierarchical triple (Chou \& Grindlay 2001; Zdziarski et al.\ 2007a). 

Then, the properties of the orbital modulation may depend on the superorbital phase. We have discussed two specific examples. In Cyg X-1, the orbital modulation is strongest at the lowest observed flux, which appears to be due to the presence of an accretion bulge at the outer edge of the precessing disc. The bulge goes into the line of sight when the disc is most edge-on, which then corresponds to the lowest observed flux. An opposite effect is observed in 4U 1820--303. Here, the orbital modulation is weakest at the lowest observed flux. The reason for that appears that the size of the obscuring bulge increases with the increasing $\dot M$, periodically changing in that system (unlike Cyg X-1, where most likely there is no periodic $\dot M$ modulation).

\vspace{1pc}
\noindent 
AAZ has been supported by the Polish MNiSW grant NN203065933 (2007--2010) and the Polish Astroparticle Network 621/E-78/SN-0068/2007. This study has been partially supported by the Academy of Finland grants 110792 and 112986, and the International Space Science Institute (Bern). AI has been supported by the Finnish Graduate School in Astronomy and Space Physics, the V\"ais\"ala Foundation and the Russian President's program for support of leading science schools (grant NSH-4224.2008.2). LW has been supported by the Alexander von Humboldt Foundation's Sofja Kovalevskaja Programme funded by the German Federal Ministry of Education and Research.

\section*{References}

\re
Anderson S. F., Margon B., Deutsch E. W., Downes R. A., Allen R. G., 1997, ApJ, 482, L69

\re
Arons J., King I.~R., 1993, ApJ, 413, L121 

\re
Ba{\l}uci\'nska-Church M., et al., 2000, MNRAS, 311, 861

\re
Basko M.~M., 1978, ApJ, 223, 268 

\re
Bednarek W., 1997, A\&A, 322, 523

\re
Bloser P.~F., et al., 2000, ApJ, 542, 1000

\re
Boroson B., Kallman T., Blondin J. M., Owen M. P., 2001, ApJ, 550, 919

\re
Brocksopp C., et al., 1999a, MNRAS, 309, 1063

\re
Brocksopp C., Tarasov A. E., Lyuty V. M., Roche O., 1999b, A\&A, 343, 861

\re
Chou Y., Grindlay J.~E., 2001, ApJ, 563, 934

\re
Coe M. J., 2000, ASPC, 214, 656

\re
Cropper M., et al., 1998, MNRAS, 293, L57 

\re
Done C., Gierli{\'n}ski M., Kubota A., 2007, A\&ARv, 15, 1 

\re
Dubus G., 2006, A\&A, 451, 9

\re
Farrell S.~A., Sood R.~K., O'Neill P.~M., Dieters S., 2008, MNRAS, in press

\re
Gies D. R., Bolton C. T., 1986, ApJ, 304, 389

\re
Gregory P. C., 2002, ApJ, 575, 427 

\re
Gregory P. C., Peracaula M., Taylor A. R., 1999, ApJ, 520, 376 

\re
Heinz S., Nowak M.~A., 2001, MNRAS, 320, 249

\re
Hellier C., Mason K. O., 1989, MNRAS, 239, 715

\re
Homer L., et al., 2001, MNRAS, 322, 827

\re
Ibragimov A., Zdziarski A. A., Poutanen J., 2007, MNRAS, 381, 723

\re
Israel G.~L., et al., 1999, A\&A, 349, L1 

\re
Israel G.~L., et al., 2002, A\&A, 386, L13

\re
Karitskaya E. A., 2001, Astron. Rep., 45, 350 

\re
Katz J. I., 1973, Nat.\ Phys.\ Sci., 246, 87

\re
Katz J. I., 1980, ApJ, 236, L127 

\re
Kozai Y., 1962, AJ, 67, 591

\re
Lachowicz P., et al., 2006, MNRAS, 368, 1025

\re
Larwood J., 1998, MNRAS, 299, L32

\re
Negueruela I., 2004, preprint (astro-ph/0411335)

\re
Ogilvie G.~I., Dubus G., 2001, MNRAS, 320, 485 

\re
\"Ozdemir S., Demircan O., 2001, Ap\&SS, 278, 319

\re
Parkinson P.~M.~S. et al., 2003, ApJ, 595, 333 

\re
Parmar A.~N., White N.~E., 1988, Mem.\ S. A. It., 59, 147 

\re
Postnov K.~A., Shakura N.~I., 1987, Sov.\ Astr.\ Lett., 13, 122 (PAZh, 13, 300)

\re
Poutanen J., Zdziarski A.~A., Ibragimov A., 2008, MNRAS, in press (PZI08)

\re
Raichur H., Paul B., 2008, MNRAS, 387, 439

\re
Ramsay G., Hakala P., Cropper M., 2002, MNRAS, 332, L7 

\re
Scott D.~M., Leahy D.~A., 1999, ApJ, 510, 974 

\re
Sood R., Farrell S., O'Neill P., Dieters S., 2007, AdSpR, 40, 1528 

\re
Stella L., Priedhorsky W., White N. E., 1987, ApJ, 312, L17

\re
Szostek A., Zdziarski A. A., 2007, MNRAS, 375, 793

\re
Wen L., Cui W., Levine A. M., Bradt H. V., 1999, ApJ, 525, 968

\re
Wen L., Levine A.~M., Corbet R.~H.~D., Bradt H.~V., 2006, ApJS, 163, 372

\re
White N.~E., Holt S.~S., 1982, ApJ, 257, 318 

\re
White N. E., Swank J. H., 1982, ApJ, 253, L61
 
\re
Zdziarski A. A., Gierli\'nski M., 2004, Progr.\ Theor.\ Phys.\ Suppl., 155, 99

\re
Zdziarski A. A., et al., 2004, MNRAS, 351, 791

\re
Zdziarski A.~A., Wen L., Gierli{\'n}ski M., 2007a, MNRAS, 377, 1006

\re
Zdziarski A.~A., Gierli{\'n}ski M., Wen L., Kostrzewa, Z., 2007b, MNRAS, 377, 1017 (Z07)

\re
Zdziarski A. A., Neronov A., Chernyakova M., 2008, preprint (arXiv:0802.1174)

\label{last}

\end{document}